\documentclass{PoS}

\title{Flavor Questions for the LHC}

\ShortTitle{Flavor Questions for the LHC}

\author{\speaker{Jonathan Rosner}\\
        Enrico Fermi Institute and Department of Physics\\
	University of Chicago\\
        E-mail: \email{rosner@hep.uchicago.edu}}

\abstract{The physics underlying quark and lepton masses and mixings (the
``flavor problem'')  is the least well understood aspect of
the Standard Model.  Some questions of flavor physics, and ways in
which the LHC can help shed light on this problem, are described.}

\FullConference{Flavor Physics and CP Violation 2009\\
		 May 27-June 1 2009\\
		 Lake Placid, NY, USA }
\def \beq{\begin{equation}}

\def \eeq{\end{equation}}
\def \lra{\leftrightarrow}
\def \ob{\overline{B}^0}
\begin{document}

\section{Introduction}

\begin{figure}
\begin{center}
\includegraphics[width=0.5\textwidth]{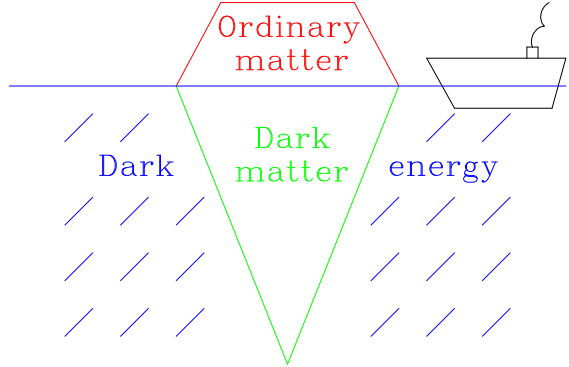}
\end{center}
\caption{Illustration of our relation to ordinary matter.
\label{fig:iceberg}}
\end{figure}

Flavor is the least well understood aspect of the Standard Model.  Ordinary
matter makes up 4.6\% of the known energy density of Universe, while dark
matter comprises another 23\% \cite{Komatsu:2008hk}.  We have little clue as to
its nature.  Dark energy accounts for the remaining 72\%;  we know even less
about it.  Ordinary quarks and leptons thus represent just the tip of a very
big iceberg (see Fig.\ \ref{fig:iceberg}).  We need to understand the rest of
the iceberg (and the sea in which it swims) in order to understand the
pattern underlying the known forms of matter.

\section{Quark patterns}

We are accumulating very precise information about the pattern of quark masses
and couplings.  If we regard the weak charge-changing transitions $u \lra d$,
$c \lra s$, and $t \lra b$ as of relative strength ${\cal O}(1)$, then the
transitions $u \lra s$ and $c \lra d$ are of order $\lambda \sim 0.23$; the
transitions $c \lra b$ and $t \lra s$ are of order $\lambda^2 \sim 0.04$, and
the transitions $u \lra b$ and $t \lra d$ are of order $\lambda^3 \sim 0.01$
or less.  This information is encoded in the unitary Cabibbo-Kobayashi-Maskawa
(CKM) matrix \cite{Cab,KM}, whose invariant phase describes CP violation.  The
CKM matrix arises from the same (unknown) physics giving rise to the pattern
of quark masses.  A related pattern arises for the leptons, which differ by
having very small neutrino masses and large mixings.

\begin{figure}
\begin{center}
\mbox{\includegraphics[width=0.58\textwidth]{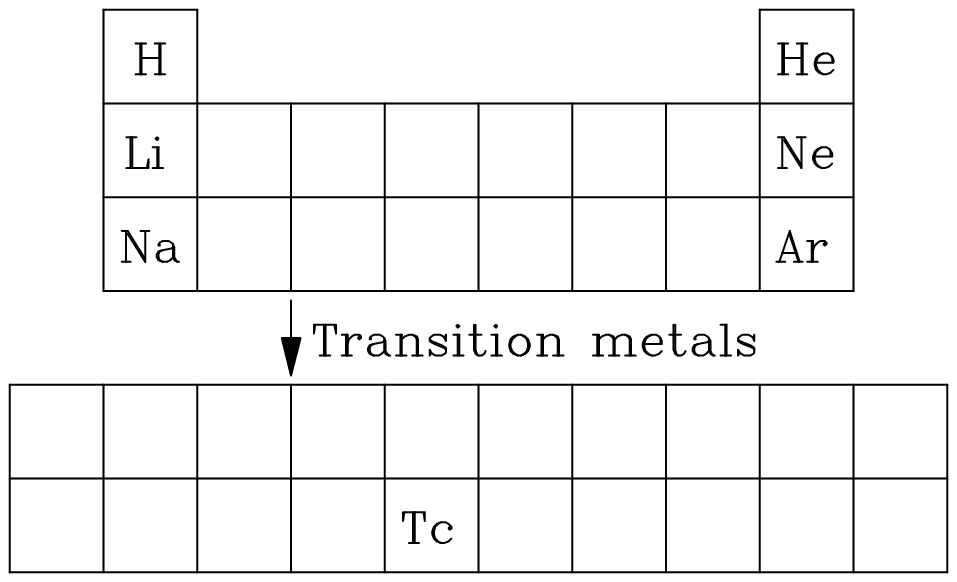} \hskip 0.2in
      \includegraphics[width=0.36\textwidth]{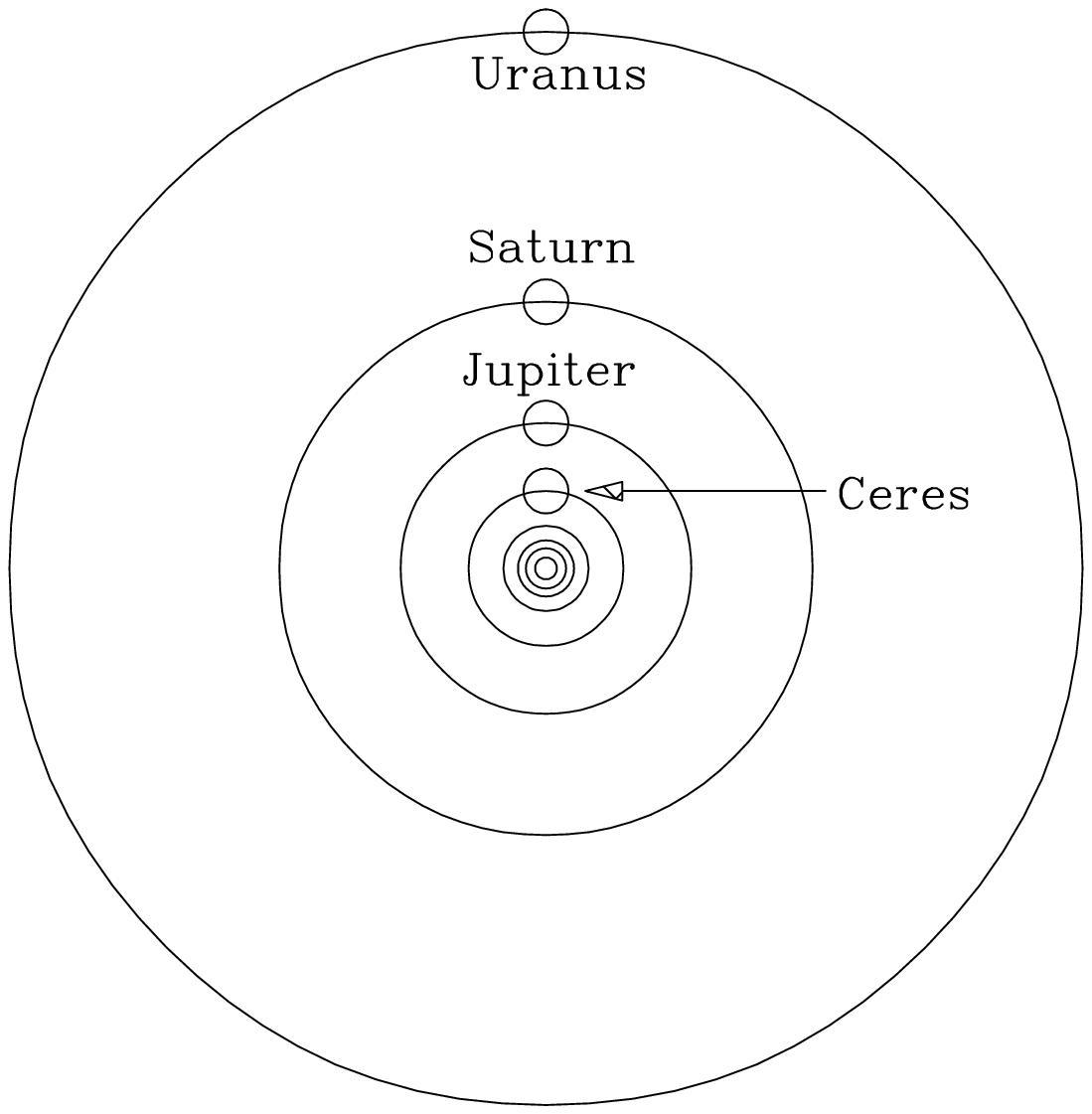}}
\end{center}
\caption{Patterns of the elements (left) and planets (right).
\label{fig:pat}}
\end{figure}

What kind of physics is giving rise to this pattern?  It is likely we will
understand it much more fully if we know how much of the pattern we are
already seeing.  Two familiar examples, illustrated in Fig.\ \ref{fig:pat},
give conflicting prospects for understanding the flavor problem.

In the periodic table of the elements, the {\it variations} of the pattern are
the key to its comprehension.  Each element has a different nuclear charge;
the electron shell structure governs chemistry.  Through this pattern, the
existence of the element Technetium was predicted.

The orbits of the planets (out to Uranus) obey the approximate relation
(Titius/Bode law) $a$(AU)$ = 0.4 + 0.3k$, where $k = 0, 1, 2, 4, 8$.  This
rule predicted the orbits of the large asteroid Ceres and the planet Uranus.
However, it failed to predict the orbit of Neptune.  Pluto is approximately
where Neptune should have been; other dwarf planets don't fit; and there is no
dynamical explanation for the rule.  Simulations can give similar relations,
in analogy to ``anarchy'' \cite{an} in models of quark and lepton masses.

Will the pattern of quark and lepton masses reflect some underlying structure,
as in the periodic table, or essential anarchy, as in planetary orbits?  We are
likely to know much more once the nature of dark matter is revealed.

\begin{figure}
\begin{center}
\mbox{\hskip 0.2in \includegraphics[width=0.96\textwidth]{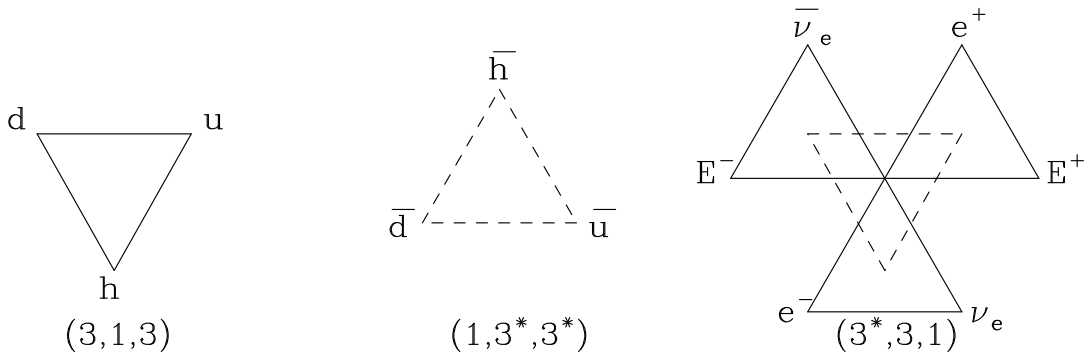}}
\end{center}
\caption{Members of the 27-plet of E$_6$, depicted in terms of their
quantum numbers in SU(3)$_{\rm L}$ (solid triangles) and SU(3)$_{\rm R}$
(dashed triangles).  From left to right:  quarks, antiquarks, and leptons.
\label{fig:e6f}}
\end{figure}

Examples of extensions of the Standard Model include a fourth family of quarks
and leptons, extended grand unified theories (GUTs), and Kaluza-Klein
excitations.  To take the example of GUTs, in SU(5) the representations $5^* +
10$ account for all known left-handed quarks and leptons in a family, while
these are combined into one 16-dimensional spinor of SO(10) with the addition
of a left-handed antineutrino (presumably with large Majorana mass).  In the
exceptional group E$_{\rm 6}$ which contains SO(10) as a subgroup, the
fundamental 27-dimensional representation involves adding an SO(10) 10-plet
and singlet to the known 16-plet.  E$_{\rm 6}$ has a subgroup SU(3)$_{\rm L}
\otimes$ SU(3)$_{\rm R} \otimes$ SU(3)$_{\rm color}$, so the 27-plet may be
represented as shown in Fig.\ \ref{fig:e6f}.  

The new fermions consist of isosinglet $Q=-1/3$ quarks $h$; vector-like leptons
$E^\pm$ and their neutrinos $\nu_E,\bar \nu_E$ (center of right-hand figure);
and a new sterile neutrino $n$ (center of right-hand figure). The $h$ could mix
with $b$ and be responsible for $m_b \ll m_t$ \cite{JRmix}.  Searches at
Fermilab exclude masses up to $\sim 300$ GeV \cite{CDFh}.

If a fourth quark-lepton family exists \cite{SoniHou}, its neutrino must be
heavier than $\sim M_Z/2$,
as the invisible width of the $Z$ indicates that only three neutrinos are
light \cite{LEPEWWG}.  New particles in loops (such as fourth-family members)
will affect $W,~Z,~\gamma$ propagators and SM coupling relations.  These
effects may be described by parameters $S$ and $T$ \cite{PT}:

\beq
\frac{G_F}{\sqrt{2}} = \left(1 + \frac{\alpha S}{4 \sin^2 \theta} \right)
\frac{g^2}{8 M_W^2}~,~~~\frac{G_F \rho}{\sqrt{2}}=\frac{g^2 + g'^2}{8 M_Z^2}~,
~~~\rho \equiv 1+\alpha T~,~~~\alpha \simeq 1/129~.
\eeq

\begin{figure}
\begin{center}
\includegraphics[width=0.8\textwidth]{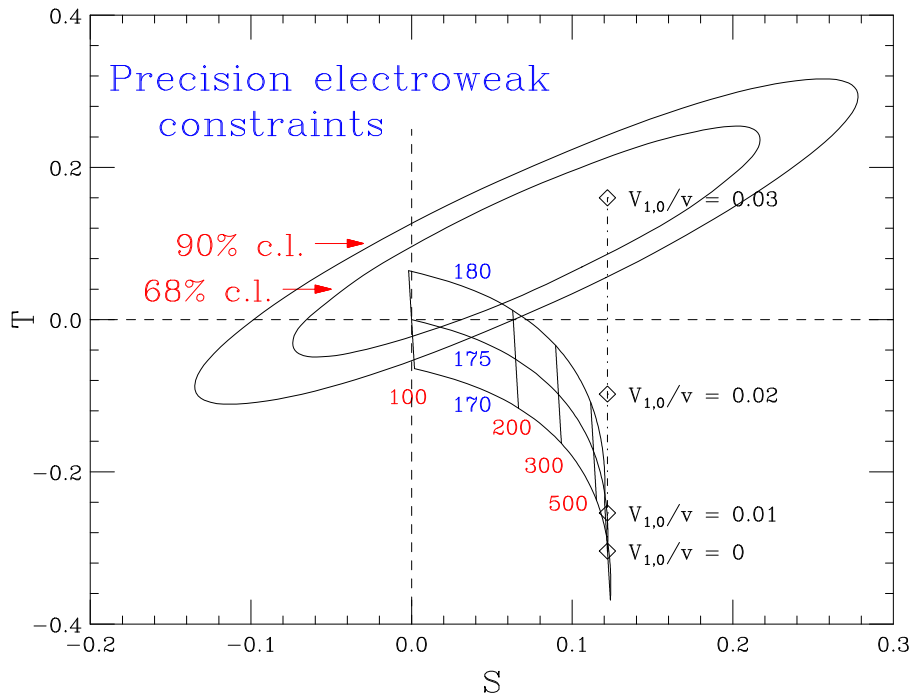}
\end{center}
\caption{$S$--$T$ plot based on precision electroweak constraints (see text).  
\label{fig:stapv}}
\end{figure}

Each new quark-lepton family contributes $\Delta S = 2/(3 \pi) \simeq 0.2$,
$\Delta T \simeq 0.4(m_{t'}^2 - m_{b'}^2)/(100~{\rm GeV})^2$.  The latter
contribution is particularly important.  In Fig.\ \ref{fig:stapv} we plot
the allowed region of $S$ and $T$ based on precision electroweak constraints
\cite{JR01}.  Also shown are predictions of the Standard Model for Higgs boson
masses of 100--500 GeV (nearly vertical contours) and top quark masses of
170, 175, and 180 GeV (curved contours).  The vertical dot-dashed line shows
the effect of a
small triplet-Higgs VEV $V_{1,0}$ (up to 0.03 of the Standard Model VEV
$v = 246$ GeV), where the subscripts denote weak isospin and weak hypercharge.
The triplet Higgs leads to $\Delta \rho = 4(V_{1,0}/v)^2$.  A large $t'$--$b'$
mass splitting behaves like a triplet Higgs, causing positive $\Delta \rho =
\alpha \Delta T$ and allowing the relaxation of the usual stringent upper
limit on the Higgs boson mass \cite{Holdom:2009}.

\section{CKM matrix parameters}

In the parametrization suggested by Wolfenstein \cite{Wolf:1983},

\beq
V = \left[ \begin{array}{c c c}
V_{ud} & V_{us} & V_{ub} \\
V_{cd} & V_{cs} & V_{cb} \\
V_{td} & V_{ts} & V_{tb} \end{array} \right] \simeq \left[ \begin{array}{c c c}
1 - \frac{\lambda^2}{2} & \lambda & A \lambda^3 (\rho - i \eta) \\
-\lambda & 1 - \frac{\lambda^2}{2} & A \lambda^2 \\
A \lambda^3 (1 - \rho - i \eta) & - A \lambda^2 & 1 \end{array} \right]~,
\eeq
the parameters are known fairly accurately:  $\lambda \simeq 0.2255$, $A \simeq
0.81$, $0.14 \le \rho \le 0.18$, and $0.34 \le \eta \le 0.36$.  (Two groups
\cite{CKMfitter,UTfit} obtain slightly different parameters when fitting
observables.)  The unitarity of $V$ implies (e.g.) $V_{ud}V^*_{ub} + V_{cd} 
V^*_{cb} + V_{td} V^*_{tb} = 0$ or dividing by the middle term, $(\rho + i
\eta) + (1 - \rho - i \eta) =1$.  This generates the {\it unitarity triangle}
(UT), whose angles opposite the sides 1, $\rho + i \eta$, $1 - \rho - i \eta$
are, respectively, $\alpha$, $\beta$, and $\gamma$.  One learns its shape from
such observables as kaon CP violation (essentially constraining $\eta(1 -
\rho)$), $B$--$\bar B$ mixing (constraining $|1 - \rho - i \eta|$, given
suitable hadronic information), and charmless $B$ decays (constraining $|\rho +
i \eta|$).  Direct measurements of angles satisfy $\alpha + \beta + \gamma =
\pi$ \cite{HFAG}, with
\beq
\alpha = (89.0^{+4.4}_{-4.2})^\circ~,~~\beta = (21.0 \pm 0.9)^\circ~,~~
\gamma = (70^{+27}_{-29})^\circ~~.
\eeq
The large error on $\gamma$ highlights the importance of improving direct
measurements of it, one of the goals of LHCb.  Measurements of {\it sides}
of the UT are more constraining, as we shall now see.

\section{Mixing of strange $B$'s}

\subsection{Constraint on CKM parameters}

A $\bar B_s^0 = b \bar s$ can mix with a $B_s^0 = s \bar b$ by means of box
diagrams involving exchange of a pair of $W$ bosons and intermediate $u,c,t$
quarks.  The heavy top quarks provide the dominant constribution, so $\bar
B_s^0$--$B_s^0$ mixing is stronger than $\bar B^0$--$B^0$ mixing because
$|V_{ts}/V_{td}| \simeq 5$.  As CKM unitarity implies $|V_{ts}| \simeq |V_{cb}|
\simeq 0.041$ (and hence $|V_{ts}|$ is well-known), $B_s$--$\overline{B}_s$
mixing probes
hadron physics.  The matrix element between $B_s$ and $\overline{B}_s$ involves
a combination $f_{B_s}^2 B_{B_s}$: $f_{B_s}$ is the ``$B_s$ decay constant''
(the matrix element of $b \bar s$ operator between $B_s$ and vacuum), and
$B_{B_s} \simeq 1$ parametrizes degree to which $W$ exchange graphs dominate
mixing.  A recent prediction of lattice QCD \cite{Lattxi}, $f_{B_s}
\sqrt{B_{B_s}}/[f_B \sqrt{B_B}] = 1.258\pm0.033$, when combined with the
well-measured $B^0$--$\ob$ mixing amplitude $\Delta m_d = (0.507 \pm 0.005)$
ps$^{-1}$ and the CDF $B_s$ mixing measurement at Fermilab \cite{CDFBsmix},
$\Delta m_s=(17.77 \pm 0.10 \pm 0.07)$ ps$^{-1}$, gives $|V_{td}/V_{ts}| =
0.214 \pm 0.005$ and hence $|1 - \rho - i \eta| = 0.950 \pm 0.026$, implying
$\gamma = (72 \pm 5)^\circ$.  This is a great improvement over the value
based on $\Delta m_d$, which was subject to uncertainty in $f_B$.
The study of $B^+ \to D^0~(\bar D^0) K^+$ may improve this accuracy eventually,
with help from information on the strong phases in the $K_S \pi^+ \pi^-$ Dalitz
plot \cite{CLEOkspipi}.

\subsection{Mixing and CP violation}

In the Standard Model (SM), $B_s \to J/\psi \phi$ is expected in SM to have a
small CP asymmetry, governed by the $B_s$--$\bar B_s$ mixing phase $\phi_M =
-2 \beta_s$, where
\beq
\beta_s \equiv {\rm Arg}(-V_{ts}V^*_{tb}/V_{cs}V^*_{cb}) = \lambda^2 \eta
\simeq 0.02~{\rm with}~\lambda = 0.2255 \pm 0.0019, \eta \simeq 0.36~.
\eeq
From angular distributions of decay products one must extract three independent
partial waves ($L=0,1,2)$ or three independent amplitudes $A_0, A_\parallel,
A_\perp$.  At the Fermilab Tevatron, both CDF \cite{Aaltonen:2007he} and D0
\cite{D0Bs}
favor a mixing phase differing from $-2 \beta_s$.  Defining $\phi_{B_s} =
\beta_s + \phi_M/2$, the HFAG average \cite{HFAG,Chandra} is $\phi_{B_s} \in
[-163,-95]^\circ,~ [-84,-17]^\circ$, $2.2 \sigma$ away from the SM.  The
width difference between CP = + and CP = -- mass eigenstates, $\Delta
\Gamma_s \simeq 0.1$ ps$^{-1}$, is compatible with SM predictions \cite{SMDG}.

A discrete ambiguity $\phi_M \to \pi - \phi_M$ is associated with uncertainty
in the strong phases $\delta_\parallel \equiv {\rm Arg} (A_\parallel A^*_0)$,
$\delta_\perp \equiv {\rm Arg} (A_\perp  A^*_0)$.  It can be eliminated by
comparison with $B^0 \to J/\psi K^{*0}$ as most contributions are similar
\cite{Gronau:2008hb}; the phases are predicted to be equal within $10^\circ$.

There are plenty of models that can accommodate a $B_s$--$\bar B_s$ mixing
phase larger than in the SM.  For examples (``littlest Higgs,'' extra
dimensions, etc.) see Ref.\ \cite{Buras:2009dy}.

\subsection{Time-dependences}

\begin{figure}
\begin{center}
\includegraphics[width=0.7\textwidth]{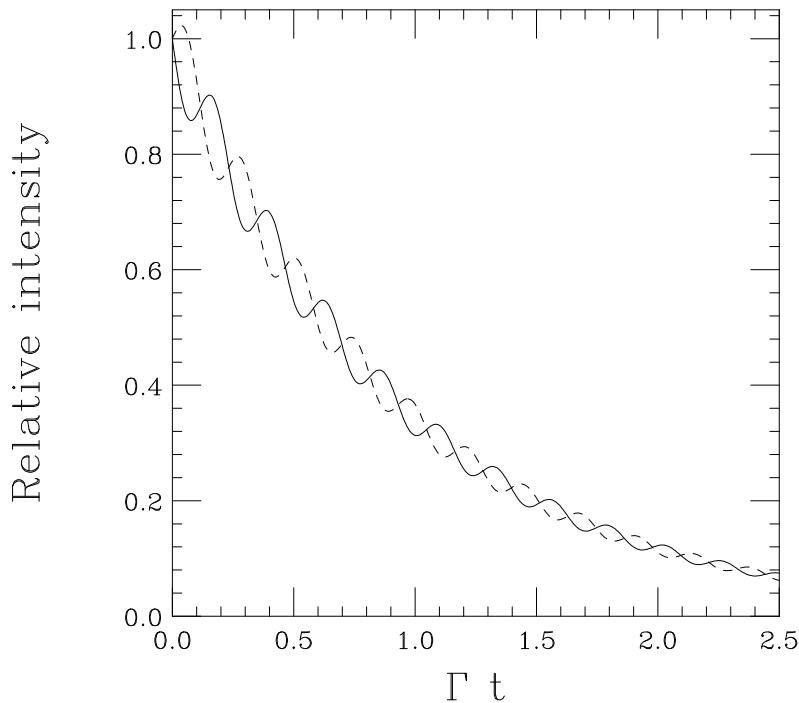}
\end{center}
\caption{Assumed time-dependence of signals for functions ${\cal T}_+$ with
tagged initial $B_s$ (solid) and $\bar B_s$ (dashed).  Based on best-fit
parameters of Ref.\ \cite{HFAG}.  Curves for ${\cal T}_-$ are
similar.
\label{fig:bf_tplus.ps}}
\end{figure}

In Ref.\ \cite{Gronau:2008hb} it was noted that the large phase claimed by CDF
and D0 for $B_s$--$\bar B_s$ mixing should lead to an explicit time-dependence
which exhibits CP violation.  Observing this will not be easy, as the flavor
oscillations are quite rapid (recall the large value of $\Delta m_s$). However,
with $\phi_M = -44^\circ$ and $\Delta \Gamma/\Gamma = 0.228$, the central
values quoted in Ref.\ \cite{HFAG}, the oscillations should be
visible, as illustrated in Fig.\ \ref{fig:bf_tplus.ps}.  Here we have defined 
functions
\beq \label{eqn:tdep}
{\cal T}_\pm e^{- \Gamma t} [\cosh(\Delta \Gamma t)/2 \mp \cos(\phi_M)
\sinh(\Delta \Gamma t)/2) \pm \eta \sin(\phi_M) \sin (\Delta m_s t)]~,
\eeq
associated with $|A_\parallel|^2$ and $|A_\perp|^2$, respectively,
which may be obtained from an angular analysis of decay products.  One
isolates CP violation by tagging at $t=0$: $\eta = \pm 1$ for tagged $(B_s,
\bar B_s)$.

The last term in Eq.\ (\ref{eqn:tdep}) contains the rapid time oscillations,
and changes sign with the sign $\eta$ of the tag.  We have assumed the
tagging parameter $\eta$ to include a dilution factor 0.11.  A plot such as
Fig.\ \ref{fig:bf_tplus.ps} would be clear evidence for non-standard
CP violation in $B_s \to J/\psi \phi$.  Such oscillations would probably be
too small to see in the Standard Model.

\section{$B \to K \pi,~\pi \pi$}

\begin{figure}[h]
\begin{center}
\includegraphics[width=0.8\textwidth]{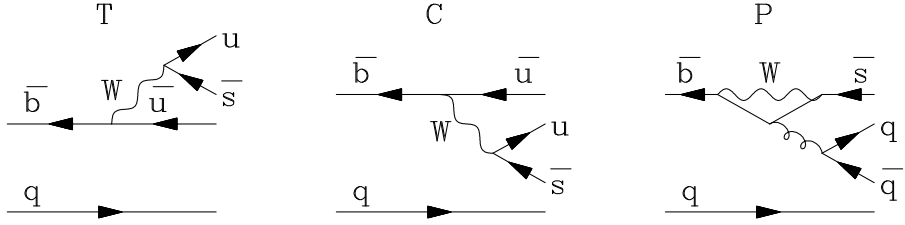}
\end{center}
\caption{Amplitudes contributing to $B \to K \pi$ decay modes.  $T$:
color-favored tree; $C$:  color-suppressed tree; $P$:  penguin.  The
annihilation graph $A$ is not shown.
\label{fig:tcp}}
\end{figure}

Some time ago it was predicted that the CP asymmetries in $B^0 \to K^+ \pi^-$
and $B^+ \to K^+ \pi^0$ would be equal if a color-suppressed amplitude
contributing to the latter process were neglected \cite{Gronau:1998ep}.
The graphical representations of various amplitudes are shown in Fig.\
\ref{fig:tcp}, and their contributions to $B \to K \pi$ processes are shown
in Table \ref{tab:bkpi}, where we have included a small ``annihilation''
($A$) contribution.  The amplitudes denoted by small letters are related
to those with large letters by the inclusion of electroweak penguin
contributions:
\beq
t \equiv T + P^C_{\rm EW}~, ~~c \equiv C + P_{\rm EW},~~p \equiv P -
(1/3)P^C_{\rm EW}~.
\eeq

\begin{table}[h]
\caption{Contributions of amplitudes to $B \to K \pi$ decay modes.
\label{tab:bkpi}}
\begin{center}
\begin{tabular}{c c c c} \hline \hline
Decay & Amplitude & BR ($10^{-6}$) & $A_{CP}$ \\ \hline
$B^0 \to K^+ \pi^-$ & $-(t + p)$ & $19.4\pm0.6$ & $-0.097\pm0.012$ \\
$B^+ \to K^+ \pi^0$ & $-(t+p+c+A)/\sqrt{2}$&$12.9\pm0.6$&$0.050\pm0.025$\\
$B^0 \to K^0 \pi^0$ & $(p-c)/\sqrt{2}$ & $9.8\pm0.6$ & $0.00\pm0.10$ \\
$B^+ \to K^0 \pi^+$ & $p+A$ & $23.1 \pm 1.0$ & $0.009\pm0.025$ \\ \hline \hline
\end{tabular}
\end{center}
\end{table}

However, the color-suppressed amplitude is {\it not} negligible.  An SU(3) fit
to $B \to (K\pi,\pi\pi)$ \cite{Chiang:2004nm} finds $|C/T| = 0.46^{+0.43}_
{-0.30}$, Arg$(C/T) = (-119 \pm 15)^\circ$.  These values have been confirmed
in a more recent analysis \cite{Li:2009wb}.  They lead to a significant
difference between the CP asymmetries in $B^0 \to K^+ \pi^-$ and $B^+ \to
K^+ \pi^0$.  So, what's the problem?  Why has this difference in CP asymmetries
been repeatedly quoted as evidence for new physics?

The debate turns on whether {\it a priori} calculations of $C$, which give
a smaller-than-observed value, can be trusted.  A large $C$ also is needed to
understand the larger-than-expected value of ${\cal B}(B^0 \to \pi^0 \pi^0) =
(1.55 \pm 0.19) \times 10^{-6}$.  The fact that no similar enhancement of $C$
appears needed in $B \to \rho \rho$ has been ascribed in Ref.\ \cite{Li:2009wb}
to a special role for pseudoscalars.  It also has been explained
\cite{Kaidalov:2007yh} in terms of rescattering:  as ${\cal B}(B \to \rho \rho)
\gg {\cal B}(B \to \pi \pi)$, the rescattering $(\rho \rho \to \pi \pi)$ is
more significant than $(\pi \pi \to \rho \rho)$, implying a greater fraction
of $C$ in $\pi \pi$
than in $\rho \rho$.  The rescattering via $\bar b \to\bar c c \bar s$ also is
a likely source of the enhanced $\bar b \to \bar s$ ``charming'' penguin.

The consistency of a unified description of $B \to K \pi$ CP asymmetries may be
tested by a robust sum rule for $A_{CP}$ which is satisfied as long as there
are no new-physics sources of a $\Delta I = 1$ amplitude \cite{Gronau:2005kz}:
\beq
\Delta(K^+ \pi^-) + \Delta(K^0 \pi^+)=2 \Delta(K^+ \pi^0) + 2\Delta(K^0
\pi^0)~,~~ \Delta(f) \equiv \Gamma(\bar B \to \bar f) - \Gamma(B \to f)~,
\eeq
which predicts $A_{CP}(B^0 \to K^0 \pi^0) = -0.148 \pm 0.044$, to be compared
with the experimental value $-0.01 \pm 0.10$.  (Furthermore, flavor SU(3)
implies a large $A_{CP}(B^0 \to \pi^0 \pi^0) \simeq 0.8$.)  The SM seems to be
able to accommodate a modestly large value of $C$; there is no need for
new-physics scenarios involving a $P_{\rm EW}$ contribution to $c = C + P_{\rm
EW}$.  The $A_{CP}$ sum rule provides a diagnostic for $\Delta I = 1$ new
physics \cite{Baek:2009hv}.  One must measure $A_{CP}(B^0 \to K^0 \pi^0)$ to
0.03 or better.

\section{Inclusive $D \to \omega X$}

CLEO's measurement of a large inclusive branching fraction ${\cal B}(D_s^+
\to \omega X) = (6.1 \pm 1.4)\%$ \cite{Dobbs:2009ni} was a surprise.  Before
this measurement, the only known $D_s$ mode involving $\omega$ was $D_s^+ \to
\pi^+ \omega$ with branching fraction ${\cal B} = (0.25 \pm 0.09\%)$
\cite{PDG}.  Now, however, CLEO has discovered a number of other $D_s$
exclusive modes involving $\eta$ \cite{Ge:2009vk}, accounting for a total of
$(5.4 \pm 1.0)\%$ of $D_s$ decays.

\begin{figure}[h]
\begin{center}
\mbox{\includegraphics[width=0.45\textwidth]{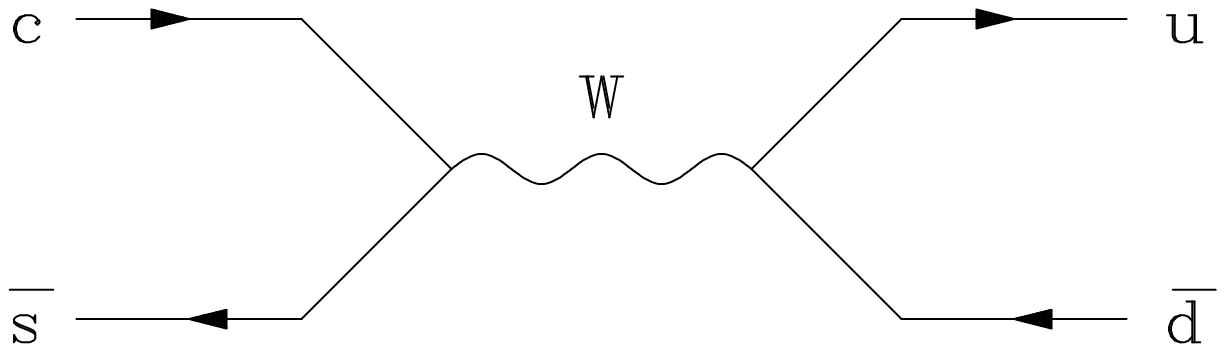} \hskip 0.2in
      \includegraphics[width=0.45\textwidth]{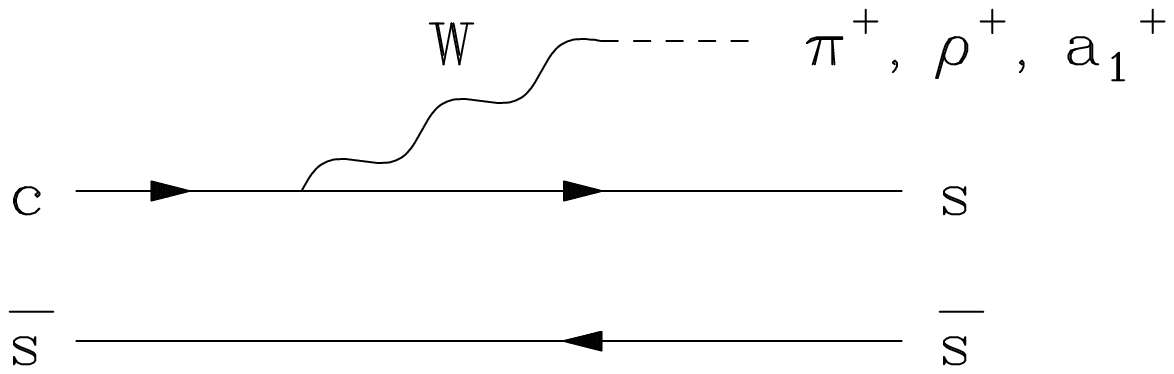}}
\end{center}
\caption{Diagrams contributing to $D_s \to \omega X$.  Left:  annihilation;
right:  color-favored tree.
\label{fig:cs}}
\end{figure}

Mechanisms for $D^+_s \to \omega X^+$ are not so obvious: one often has to
get rid of an $s \bar s$ pair.  Two candidate subprocesses are shown in
Fig.\ \ref{fig:cs}.

In the left-hand diagram, the process $D_s^+ \to ({\rm virtual}~W^+) \to u \bar
d$ is helicity-suppressed, and G-parity forbids production of the final states
$\pi^+ \omega$ and $(3 \pi)^+ \omega$.  In the right-hand diagram, the
subprocess $c \to u \bar d s$ with a spectator $s$ could give $\omega \pi^+
\eta$ \cite{Gronau:2009vt}.  One also could get $\omega(\pi^+,\rho^+,a_1^+)$
if the transition $s \bar s \to \omega$ is somehow not subject to the usual
Okubo-Zweig-Iizuka (OZI) suppression \cite{Gronau:2009mp}.  In that case
one might expect $D_s \to \omega \ell^+ \nu_\ell$ to be observable.  Helicity
suppression also seems not to be apparent in CLEO's result \cite{Athar:2008ug}
${\cal B}(D_s \to p \bar n) = (1.30 \pm 0.36^{+0.12}_{-0.16}) \times 10^{-3}$,
given a reasonable form factor for the weak current to produce $p \bar n$.

\section{Some models for new physics}

Extra $Z$ bosons arise in many extensions of the SM.  They are not guaranteed
to have flavor-diagonal couplings if SM fermions also mix with new fermions in
such extensions.  For example, GUTs based on the exceptional group E$_{\rm 6}$
have two extra $Z$ bosons $Z_\chi,Z_\psi$ (only one linear combination of
which may be relatively light) and extra isoscalar quarks with $Q = -1/3$
which can mix with $d,s,b$.

Many grand unified theories have a ${\rm SU(4)_{color} \times SU(2)_L \times
SU(2)_R}$ subgroup \cite{Pati:1973}.  SU(4)$_{\rm color}$ unifies quarks and
leptons and contains U(1)$_{\rm B-L}$ and leptoquarks; SU(2)$_{\rm R}$ has
right-handed $W$'s and a U(1)$_{\rm R}$ such that the electromagnetic charge
is $Q = I_{\rm 3L} + I_{\rm 3R} + (B-L)/2$.  Leptoquarks can contribute to
leptonic meson decays; right-handed $W$'s contribute to mixing; there are
strong constraints on $W_L$--$W_R$ box diagrams (see, e.g., Ref.\
\cite{Langacker:1989xa}.

In supersymmetry, box diagrams can change flavor unless specifically forbidden.
Electroweak-symmetry-breaking schemes (for example, littlest-Higgs models with
T-parity, technicolor, etc.) generically have flavor-changing interactions.
Theories with extra dimensions (a recent example is Ref.\
\cite{Fitzpatrick:2008zza}) can concentrate flavor violation in the top sector
(a particular target for the International Linear Collider), and can possess
Kaluza-Klein excitations at the TeV scale, accessible at the LHC.

\section{Dark matter scenarios}

Imagine a TeV-scale effective symmetry SU(3) $\otimes$ SU(2) $\otimes$ U(1)
$\otimes$ G, where the beyond-Standard-Model (BSM) group G could be SUSY with
R-parity, extra-dimensional excitations with Kaluza-Klein parity, little Higgs
models with T-parity, technicolor, or some other group.  One can classify the
types of matter very generally as shown in Table \ref{tab:types} 
\cite{Rosner:2005ec}:

\begin{table}[h]
\caption{Possible types of matter classified according to SM and BSM (G)
transformation.
\label{tab:types}}
\begin{center}
\begin{tabular}{c c c c} \hline \hline
Type of matter & Std.\ Model &    G    & Example(s) \\ \hline
Ordinary       & Non-singlet & Singlet & Quarks, leptons \\
Mixed          & Non-singlet & Non-singlet & Superpartners \\
Shadow         & Singlet     & Non-singlet & $E_8'$ of E$_8 \otimes$
E$_8'$ \\ \hline \hline
\end{tabular}
\end{center}
\end{table}

Ordinary matter could be singlets under $G$ even if its subconstituents were
non-singlets (e.g., in composite-Higgs models).  Loops could involve
$G$-nonsinglets.  Many dark matter scenarios involve mixed matter, such as
superpartners or particles with odd KK- or T-parity.  Flavor-changing loops
can occur.  Mixed-matter scenarios may be different if G is more general
than a ``parity.'' Shadow matter may not interact with ordinary matter {\it at
all} except gravitationally.

\section{Hidden sector in loops}

\begin{figure}
\begin{center}
\mbox{\includegraphics[width=0.31\textwidth]{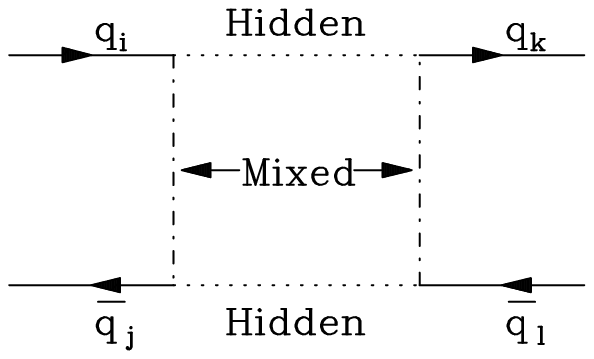}
      \includegraphics[width=0.35\textwidth]{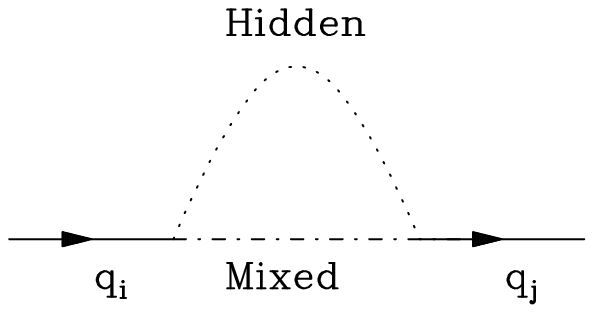}
      \includegraphics[width=0.24\textwidth]{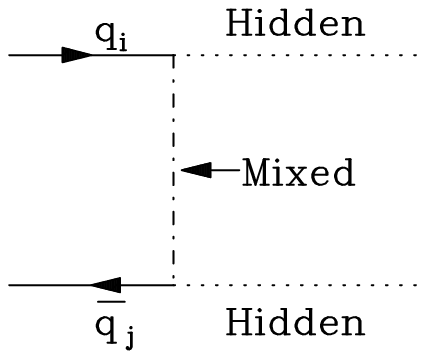}}
\end{center}
\caption{Hidden sector interacting with ordinary
matter.  Left:  box diagram; center: penguin diagram; right: production of a
pair of hidden-sector particles through exchange of a mixed state.
\label{fig:hid}}
\end{figure}

Manifestations of a hidden sector interacting with ordinary matter are shown
in Fig.\ \ref{fig:hid}.
Mixed particles must have the same SU(3) $\otimes$ SU(2) $\otimes$ U(1) quantum
numbers as the quarks to which they couple, but off-diagonal flavor couplings
are allowed.  Flavor-diagonal couplings still can affect such quantities as
the muon anomalous moment $a_\mu$, which has been shown particularly sensitive
to new physics in some supersymmetry scenarios.  For a coupling ${\cal O}
(\alpha)$, the mass scale to explain the current $3 \sigma$ discrepancy in
$a_\mu$ is $\sim 50$ GeV.

In a recent paper, D. McKeen \cite{McKeen:2009rm} suggests
looking for light ``hidden'' states in quarkonium decay.  For example,
one can look for a light dark matter candidate $X$ in $\Upsilon(2S) \to \gamma
\chi_{b0} \to \gamma X X$.  This is one manifestation of the ``WIMPless
Dark Matter'' scenario of Ref.\ \cite{Feng:2008ya}.

\section{Some LHCb topics}

The LHCb experiment will provide a unique window to $B_s$ decays, through (e.g.)
(1) Better $J/\psi \phi$ studies, with explicit time dependence plots.  (2)
$B_s \to J/\psi \eta$:  although ${\cal B}$ is less (1/3 of that for $J/\psi
\phi$), no helicity analysis is needed.  (3) $B_s \to J/\psi f_0$: L. Zhang, in
a poster at this Conference, estimates ${\cal B}(B_s \to J/\psi f_0 (\to \pi^+
\pi^-))/{\cal B}(B_s \to J/\psi \phi (\to K^+ K^-)) = (42 \pm 11)\%$.
(4) A CP analysis involving $A(B_s \to D_s^+ K^-) \sim V^*_{ub} V_{cs}$;
$A(\bar B_s \to D_s^+ K^-) \sim V^*_{us}V_{cb}$.
(5) Comparisons of $(B,~B_s) \to (\pi \pi, K \pi)$ \cite{Fleischer:2000,
Gronau:2000}, yielding independent estimates of $\gamma$.
(6) Many tests of flavor SU(3) by comparison with $B$ decays.

The hidden valley scenario \cite{HV} suggests an energy threshold (if we are
lucky, the TeV scale) for the production of new matter; some may end up in new
light (few GeV?) states.  The LHCb Collaboration is aware of these
possibilities, having discussed the examples of a 3 TeV $Z'$, a 35 GeV
``v-pion,'' and a SM Higgs devaying to a pair of v-pions \cite{KS}.

Charm studies at LHCb will explore virgin territory because of the large
production cross sections and small Standard Model CP violation.  One will
be able to probe loop and penguin diagrams involving the mixed and hidden
sectors with unprecendented sensitivity.

\section{Looking forward}

Belle and the Fermilab Tevatron are still running; BaBar and CLEO are analyzing
a rich trove of data.  CLEO is capable of searching for light scalars or
pseudoscalars in bottomonium decay, and the same is to be expected of the
$B$ factories.  In the nearest future we see results from BESIII \cite{Roy}
and from LHCb whenever the
LHC begins operation, and some $b$ physics capabilities at ATLAS and CMS.
Questions include many on the strange $B$ system, e.g., pinning down the mixing
and/or the CP-violating phase in the $B_s$--$\bar B_s$ system.

Other LHCb questions include: (a) flavor symmetry and departures from it in
$B_s$ decays, to check schemes seeking to calculate
strong-interaction properties (e.g., non-factorizable amplitudes); (b) effects
of any new sector on loops and direct production of new particles.

The KEK-B/Belle upgrade will initially provide a data sample of 10 ab$^{-1}$
and eventually $> 5$ times that; super-B more.  A simple motivation
for these machines is that anything studied
previously with single-$B$ decays now can be studied with double-tagged events
if tagging efficiency approaches 1\%.  Going further in $e^+ e^-$ collisions,
we will hope for an ILC to explore the Higgs, SUSY, and top sectors.

Finally, present experience with $B$ decays tells us that a rich program of
understanding strong-interaction and nonperturbative effects will be needed to
complement searches for rare processes in order to interpret apparent
departures from the SM as genuine signs of new physics.

\section*{Acknowledgments}

I would like to thank B. Bhattacharya, C.-W. Chiang, M. Gronau, M. Karliner,
D. McKeen, B. Keren-Zur, H. Lipkin, D. Pirjol, A. Thalapillil, and my
colleagues on the CLEO Collaboration for the opportunity to work together,
and the organizers of FPCP2009 for a truly enjoyable and
informative conference in a superb setting.  This work was supported in part
by the United States Department of Energy under Grant No.\ DE-FG02090ER40560.

\end{document}